\documentstyle[preprint,aps]{revtex}
\tightenlines

\begin{document}

\newcommand{\arcsinh}{\mathrm{arcsinh}}

\title{ Computation of the phase induced by non-newtonian gravitational potentials
in atom interferometry}

\author{R. Mathevet, R. Delhuille and C. Rizzo}

\address{Laboratoire Collisions Agr\'egats R\'eactivit\'e-IRSAMC
\\ Universit\'e Paul Sabatier and CNRS UMR 5589
\\ 118, Route de Narbonne 31062 Toulouse Cedex, France
\\ e-mail:~{\tt carlo.rizzo@irsamc.ups-tlse.fr}}

\maketitle

\begin{abstract}
In this letter we present a computation of the phase induced by
test masses of different geometry, in the framework of
non-newtonian gravitation, on an ideal separated arms atom
interferometer. We deduce the related limits on the non-newtonian
gravitational strength in the sub-millimeter region for the
potential range. These limits would be comparable with the best
existing experimental limits but with the advantage of using a
microscopic probe.

\end{abstract}

\section{Introduction}

In this first section we will recall briefly the main lines of the
theoretical motivations of this kind of calculations and their
interest and the present state-of-the-art of experiments in this
field will be summarized. For a more detailed description of both
theoretical and experimental point of view the reader is referred
to \cite{KF} and \cite{LCP}. The basics of atom interferometry
will be then presented and, after some geometrical considerations,
the signal for different kinds of potential will be obtained.
Finally, we will derive the limits on the non-newtonian
gravitational strength that could be obtained by an ideal
experiment and we will compare the results to present limits.

In the attempt to unify the description of all known forces, two
types of extensions of newtonian gravity are usually made
\cite{KF}. First, one can postulate that a new force mediated by
massive scalar bosons exists and that it gives rise to a Yukawa
potential so that the total gravitational potential between
point-like particles of mass $m_1$ and $m_2$ separated by a
distance $r$ reads:

\begin{equation}
\label{yukawa} V_Y(r)=-\frac{G m_1 m_2}{r}\,(1+
\alpha_Y\exp^{-r/\lambda_Y})
\end{equation}

In the preceding expression $G$ denotes the gravitational constant
and the first term is the usual newtonian potential. The
correction term is thus characterized by its dimensionless
strength $\alpha_Y$ and its range $\lambda_Y$ which is essentially
inversely proportional to the boson mass. A peculiar feature of
such Yukawa potential is that the exponential factor saturates
when the mutual distance $r$ tends to zero: if its strength
$\alpha_Y$ is small compared to $1$ then the Yukawa correction
remains always a small correction to newtonian gravity which is
itself a very weak interaction compared to the other known forces.

On the other hand, some recent theoretical developments account
for such weakness supposing that there exists $n$ extra dimensions
in which only gravity propagates so that the main part of it is
{\it lost} for the usual 3D-space \cite{ADD}. The extra dimensions
are compact so that their existence can be felt only to mutual
distances smaller to a compactification radius $\lambda_n$: for
distances much greater than the compactification radius then the
reminder of the gravitation interaction in the extra dimensions
manifests itself as a Yukawa-like potential with a range of the
order of the compactification radius $\lambda_n$ and a strength
$\alpha_n$ of a few units. It thus quickly decreases leaving just
the weak long range usual inverse square law interaction. On the
contrary, at short mutual distances, it follows from Gauss theorem
in $3+n$ spatial dimensions that the gravitational interaction
$V_n(r)$ behaves as $r^{-(1+n)}$ so that it conveniently reads:

\begin{equation}
\label{extradim} V_n(r)=-\frac{G m_1 m_2}{r}\, \alpha_n \left(
\frac{\lambda_n}{r} \right)^n \,\, for \,\, r\ll\lambda_n
\end{equation}

It should be noticed that this potential does not saturates as the
Yukawa potential at zero distances and "produces dramatic
deviation from newtonian gravity" \cite{KF}. The model \cite{ADD}
predict a compactification radius $\lambda_2 \sim
10^{-4}-10^{-3}m$ which make it a matter of an experimental test.

Up to now, there is experimental evidence that the inverse square
law for gravitational interactions is well tested for separations
$r$ between $10^{-2}m$ and $10^{15}m$ where $\alpha$ is known to
be smaller than $10^{-4}$. A detailed description of the
experimental works can be found in reference \cite{LCP}. The best
limits on $\alpha$ in the region around $\lambda\sim10^{-4}m$ have
been very recently provided by the experiment described in
reference \cite{HSH}. This experiment relies on a torque pendulum,
the motion of which would be perturbed in a specific way by
non-newtonian potentials from a test mass. Separation between
$200\mu m$ and $10 mm$ have been achieved with a null result. The
present day constraint, therefore, for $\lambda \sim 10^{-4}m$ is
$\alpha \leq 100$ \cite{HSH}.

Up to now all the performed experiments share common features from
technical constraints and scaling considerations. In particular,
they are macroscopic in the sense that both the probe and test
masses dimensions are in the centimeter range.

In the following, we will show that atom interferometry, that uses
a microscopic probe, could explore different geometries and
distances ranges. So let us now turn to the basics of atom
interferometry.

\section{atom interferometry}

Atom interferometry aims at making interferences with de Broglie
waves associated to the external motion of massive particles. It
was born in the late 80's when mechanical effects of light were
extensively studied which lead the principal contributor, W. D.
Philips, S. Chu and C. Cohen Tannoudji to Nobel price in 1997. The
main difference with optical interferometry lies in the fact that
the particle used (atoms or molecules) possess internal degree of
freedom and mass. That make the interferometer sensitive to
external fields and inertial effects we are concerned with. For an
introductory review the reader is referred to \cite{BMR} and for
more details to \cite{Ber}. As we want to probe distance dependent
potential we will concentrate on an idealized separated arms
interferometer as outlined in Fig.~\ref{interferometer}. An atom
beam is emitted from a source $S$ and passes through a three
identical beam splitters $G_{1,\;2,\;3}$ separated by a distance
$L$. At each beam splitter, each incoming beam is coherently
divided in two parts separated by an angle $\theta$ (labels
$A,\,O,\,O',\,B$ on Fig.~\ref{interferometer}). Only one of the
closed paths is shown on Fig.~\ref{interferometer} at the end of
which the interference pattern is recorded on a detector $D$. The
detection scheme obviously depends on the particles used such as
alkaline ($Li$, $Na$, $Rb$, $Cs$) atoms or molecules,
earth-alkaline metastable atoms ($Mg^*$, $Ca^*$), metastable rare
gas atoms ($He^*$, $Ar^*$, $Ne^*$), molecules ($I_2$, $Na_2$,
$C_{60}$).

In the same way, the actual shape of the interferometer depends on
the very technique used to construct the beam splitters. The
example chosen above corresponds to grating interferometers:
atoms, whose de Broglie wavelength is $\lambda_{dB}$, interacting
with a modulated perturbation of period $\Lambda$, are diffracted
at a typical diffraction angle $\theta\sim
\frac{\lambda_{dB}}{\Lambda}$. Such a perturbation can be for
example a material grating or a laser standing wave.

In ideal conditions, the signal recorded by the detector is then
$I(\delta\phi)=I_0\cos\delta\phi$ where $I_0$ equals to the atom
flux times the interaction time and $\delta\phi$ is the phase
difference between the two arms.

In the following, to evaluate the phase difference induced by
gravitational interactions, we will consider a semi-classical
situation in which the atom velocity $v$ and mass $M$ are high
enough so that its de Broglie wavelength
$\lambda_{dB}=\frac{h}{Mv}$ is small compared to any relevant
length characterizing the potential $V({\mathbf r},t)$ existing in
between the beam splitters. This potential will also be considered
as a perturbation of the free motion of the particles which thus
fly in straight line between the source, beam splitters and
detector. Then, the phase shift accumulated by the particle with
respect to the unperturbed case can be written as
\cite{FH},\cite{SC}:

\begin{equation}
\label{phaser} \phi=\frac{1}{\hbar}\int_{\Gamma_{cl.}}\frac{dr}{v}
V({\mathbf r},t({\mathbf r}))
\end{equation}

where $\Gamma_{cl.}$ is the classical unperturbed path of the
particle. The phase difference simply reads:

\begin{equation}
\label{diffdephase}
\delta\phi=\frac{1}{\hbar}\int\limits_{\scriptscriptstyle
AO'B}\frac{dr}{v}V({\mathbf r},t({\mathbf r}))-
\frac{1}{\hbar}\int\limits_{\scriptscriptstyle
AOB}\frac{dr}{v}V({\mathbf r},t({\mathbf
r}))=\oint\limits_{\scriptscriptstyle
AO'BOA}\frac{dr}{v}V({\mathbf r},t({\mathbf r}))
\end{equation}

To effectively compute this phase difference and thus the expected
signal one has specify the geometry of the test mass to evaluate
the generated gravitation potential.

\section{geometry considerations and phase calculations}

As explained in reference \cite{LCP}, one should use small size
test mass to get as near as possible and thus increase the
relative sensitivity to non-newtonian potentials. Nevertheless, as
we investigate a totally different experimental technique, new
possibilities are opened. Obviously, a linear ($1D$) test mass
parallel to the atomic trajectory is preferable to a point-like
($0D$) one as it increases the interacting time with the probe.
One the other hand the use of a plane ($2D$) test mass is
unfavourable for two reasons \cite{newton2D}. First, as the
potentials of interest are of short range, only a stripe of width
$\sim\lambda$ parallel to the atomic trajectory will contribute
significantly to the interactions so that the $1D$ case give a
reasonable approximation. Second, an atom near a surface is
subjected to the Van der Waals potential that will screen the
other ones. On the contrary, one might expect that the
electrostatic image of the atom by a 1D-wire is weaker than by a
2D-plane leading to a smaller Van der Waals interaction that thus
will be neglected in the following. To be exhaustive, one should
go into the higher dimensions cases. The $3D$ case of typical size
$R$ is also unfavourable because only a volume $R\lambda^2$ has to
be taken into account. However, as suggested in Eq.~\ref{phaser},
the use of time-dependent potentials (e.g. an oscillating test
mass) open the way to fruitful phase sensitive detection
techniques.

We thus chose for the calculations an idealized wire test mass of
linear density $\mu$, of length $L$, located between the two first
beam splitters at a distance $d$ from the lower partial beam and
parallel to it (see Fig.~\ref{interferometer}). Let us take $G_2$
as the $x$-axis of a reference frame and its intersection with the
$z$-axis as the origin. Assuming a Yukawa extra interaction, the
potential at the position $(x,z)$ is:

\begin{equation}
\label{yukawa1D0} V_Y^{\scriptscriptstyle
1D}(x,z)=\int_{-L-z}^{-z}-\frac{\alpha_Y\,GM}{\sqrt{(d+x)^2+u^2}}\,\mu\
du\;\exp(-\frac{1}{\lambda_Y}\sqrt{(d+x)^2+u^2})
\end{equation}

In the preceding expression, $|d+x|$ is of the order of
$\lambda_Y$, thus in the sub-millimeter range,  whereas $L$ is the
length of the test mass which is related to the length of the
apparatus which is commonly in the meter range. So, except for
negligible fringe effects, the limits of the integral may be
rejected to infinity as the exponential term to be integrated
quickly decreases. Then, with the variable changes
$v=\frac{u}{|d+x|}$ and $v=\sinh w$:

\begin{equation}
\label{yukawa1D} V_Y^{\scriptscriptstyle
1D}(x,z)\approx-\alpha_Y\,GM\mu\,\int_{-\infty}^{+\infty}\,\
dw\;\exp(-\frac{|d+x|}{\lambda_Y}\cosh
w)=-\alpha_Y\,GM\mu\,2K_0(\frac{|d+x|}{\lambda_Y})
\end{equation}

where $K_0$ is the modified Bessel function of first kind
\cite{Grad}.

Between the two first gratings ($z<0$), the two partial beams can
be parameterized as $x_2(z)=0$ and $x_1(z)=e(1+\frac{z}{L})$. So
the phase delay accumulated along the lower path simply reads
$\phi_2=\frac{GMm}{\hbar v}\,2\alpha_YK_0(\frac{d}{\lambda_Y})$,
where $m=L\mu$ is simply the test mass. The phase delay
accumulated on the upper path $\phi_1$ contains the mean value of
the $K_0$ function which has a rather complicated analytical
expression involving hypergeometric and other Bessel functions.
For sake of simplicity, we will consider only the limiting case
where the beam separation $e$ is much greater than the interaction
range $\lambda$. Then the potential is practically negligible
everywhere along the upper path and so for $\phi_1$. The phase
difference is then simply $\delta\phi=\phi_2-0$:

\begin{equation}
\label{deltaphiinfiniyukawa1D} \delta\phi_Y^{\scriptscriptstyle
1D}=\frac{GMm}{\hbar
v}\,2\alpha_YK_0(\frac{d}{\lambda_Y})\,\,if\,\,e\gg\lambda_Y
\end{equation}

>From Ref. \cite{AS} we get the following expansions: $
\delta\phi_Y^{\scriptscriptstyle 1D}\sim
\sqrt{\frac{\lambda}{d}}e^{-\frac{d}{\lambda}}$ as
$\frac{\lambda}{d}\ll 1$ and $\delta\phi_Y^{\scriptscriptstyle
1D}\sim \ln{\frac{\lambda}{d}}$ as $\frac{\lambda}{d}\gg 1$

The same kind of calculations can be done for the case of extra
dimensions. Some difficulties arise because the law we have
(Eq.~\ref{extradim}) is valid only for distances shorter than the
compactification radius $\lambda_n$. For longer distances, the
potential decreases exponentially fast in a Yukawa fashion
\cite{KF} and should give a negligible contribution. Let's make
the assumption that $d\ll\lambda_n$. As done before we will
suppose that $e\gg\lambda$ to neglect the contribution of the
upper path. We will get an estimate of the effect on the lower
path restricting the integration range to a distance $\lambda_n$
around the position of the particle. Under these assumptions, and
neglecting fringe effects, the potential arising from $n>0$ extra
dimensions reads:

\begin{eqnarray}
\label{extradiminfini1D} V_n^{\scriptscriptstyle
1D}(x,z)&=&\int\limits_{-\lambda_n}^{+\lambda_n}-\frac{\alpha_n\,GM}
{\sqrt{(d+x)^2+u^2}}\,\mu\
du\;(\frac{\lambda_n}{\sqrt{(d+x)^2+u^2}})^n \nonumber\\
&=&-\alpha_n\,GM\mu(\frac{\lambda_n}{|d+x|})^n
\int\limits_{-\frac{\lambda_n}{|d+x|}}^{+\frac{\lambda_n}{|d+x|}}
dv\;(1+v^2)^{-\frac{1+n}{2}}\nonumber\\
&=&-2\alpha_n\,GM\mu(\frac{\lambda_n}{|d+x|})^(n+1)\,
{}_2F_1(\frac{1}{2},\frac{n+1}{2},\frac{3}{2},-(\frac{\lambda_n}{d+x})^2)
\end{eqnarray}

where ${}_2F_1$ denotes the hypergeometric function
\cite{specialcases}.

The phase difference then reads:

\begin{equation}
\label{deltaphiinfiniextradim1D} \delta\phi_n^{\scriptscriptstyle
1D}=\frac{GMm}{\hbar v}\,2\alpha_n\,(\frac{\lambda_n}{d})^{n+1}\,
{}_2F_1(\frac{1}{2},\frac{1+n}{2},\frac{3}{2},-(\frac{\lambda_n}{d})^2)
\,\,if\,\,e\gg\lambda_n\,\,and\,\,n>0
\end{equation}

and, in particular,

\begin{equation}
\label{deltaphiinfini2extradim1D} \delta\phi_2^{\scriptscriptstyle
1D}=\frac{GMm}{\hbar v}\,2\alpha_2\,(\frac{\lambda_2}{d})^3\,
\left(1+(\frac{\lambda_n}{d})^2\right)^{-\frac{1}{2}}
\end{equation}

One can show that, assuming that $m$ is a constant, in the region
$d\leq\lambda$ were the model holds that $
\delta\phi_n^{\scriptscriptstyle
1D}\sim\left(\frac{\lambda}{d}\right)^n$ when $d\ll\lambda$.

Nevertheless, the newtonian potential $n=0$ has infinite range and
thus must be treated separately. As the distance between the
particle and the wire is much smaller than its length, one can
take the infinite length of the wire limit \cite{fullnewton}. So,
from Gauss theorem, the gravitational field is ${\mathbf
g}({\mathbf r})=-2\,G\mu\,\frac{{\mathbf
r}}{r^2}=-{\mathbf\nabla}\left(2\,G\mu\,\ln\frac{r}{d}\right)$ so
that the potential is null on the lower path. The potential energy
is then:

\begin{equation}
\label{newtonwire} V_{newton}^{\scriptscriptstyle 1D}(x)=2\,GM\mu
\ln (1+\frac{x}{d})
\end{equation}

The newtonian phase difference is then:

\begin{equation}
\label{deltaphinewton} \phi_{newton}^{\scriptscriptstyle 1D}
=\frac{GMm}{\hbar v}\,2\ln(1+\frac{e}{d})
\end{equation}

With all these quantitative expressions for the different phases
we can now turn to numerical estimates for a reference atom
interferometer.

\section{Numerical results}

In all cases, the phase difference is $\frac{GMm}{\hbar v}$ times
a numerical factor which can be inferred from dimensional
considerations. The numerical factor depends on the particular
geometry used and the presumed force law. It should be noticed
that this numerical factor involves dimensionless ratios as
$\alpha$, $\frac{\lambda}{d}$ that can be great. Anyway, let us
first give a numerical estimate of the pre-factor, taking a slowed
cesium beam $v=10\;m.s^{-1}, M=2\;10^{-25}kg$, and a wire
$1m\times 100\mu m\times 100\mu m$ of density equal to $20$ (gold)
so that $m=2\;10^{-4}kg$. The pre-factor then amounts to some
$10^{-6}$. The detection limit $\delta\phi_{min}=10^{-3}rad$ is
commonly accepted for atom interferometry (see different
contributions in \cite{Ber}). It first confirms the known fact
that the metrology on the regular newtonian gravity (i.e. a
measurement of $G$) is practically impossible as even with
$d=10\mu m$, $\ln(1+\frac{e}{d})$ will amount to some units so
$\phi_{newton}^{\scriptscriptstyle 1D}$ (Eq.\ref{deltaphinewton})
will be to small to be detected.

One can anyway evaluate the upper bounds on the strength of the
non-newtonian interaction in the case of a hypothetical null
experiment based on our calculation. The results are shown in
Fig.~\ref{comparison} together with limits taken from Ref.
\cite{HSH}. They were obtained assuming that the actual wire
diameter equals the distance from the wire to the atoms $d$ so
that the $1D$ model is at the limit of acceptability. The upper
part corresponds to $d=100\mu m$ so that, for $\lambda<d$, the
wire is beyond the compactification radius for the models with
extra dimensions. The curves have then been extrapolated the curve
given by  the Yukawa potential. $d=10\mu m$ is shown in the lower
part for which such a problem do not arise since we have chosen
the same parameters range as in Ref. \cite{HSH} for sake of
comparison. In the same way, the beam separation $e\sim10^{-3}m$
with the chosen parameter set. The limits are thus overestimated
in the somehow low interest region $\lambda>10^{-3}m$.

While a real experiment based on this computation looks very
difficult for several reason, the figure clearly shows that there
is an interest in further studies of the methodology we described.
For short ranges, $\lambda<10^{-4}m$, the method appear
particularly efficient even if one uses a less sophisticated
atomic source such as a supersonic beam for which
$v\sim10^3m.s^{-1}$. In the end, the main free parameter in the
simulations is $d$ and has to be chosen with respect to the
potential model of interest and its suspected range. When $d$
decreases the different potentials increase but the test mass
scales typically as $d^2$ and vice versa. For the Yukawa case, an
optimum is numerically, and not surprisingly, found for
$d\sim\lambda$ for the typical parameters ranges we are concerned
with. On the contrary for n extra dimensions the overall phase
scales as $d^2\left(\frac{\lambda}{d}\right)^n\sim d^{2-n}$. The
case $n=2$ thus exhibits a weak dependency on the choice of the
wire diameter whereas one should use as small as possible a wire
if $n\geq 3$. On the contrary, if the $n=1$ case is under
investigation, one should use the largest acceptable wire that is
$d\sim\lambda$. These behaviors can be checked on the figure where
the limits for $n=1,\;2,\;3$ get worse, remain roughly unchanged
or improve from $d=100\mu m$ (upper graph) to $d=10\mu m$ (lower
graph).

\section{Conclusions}

The results presented here concerning non-newtonian gravitation
are based on a completely different method than the existing ones
that use macroscopic probes. Our computation show that atom
interferometry could provide limits on the strength $\alpha$ of
several extra potentials comparable with the existing ones in a
reasonable integration time, especially if they are of short range
$\lambda<10^{-4}m$. More, it can be easily extended to test a
composition dependent interaction ({\it "fifth-force"}) using, for
example, as test masses two wires of different materials or two
isotopes such as $^6Li$ and $^7Li$ for the probe atom. Using
polarized beams spin dependent potentials could be also taken into
account.

A detailed study of an actual experimental set-up is far beyond
the scope of such a letter; a rapid realization of such an
experiment looks anyway difficult because of some experimental
critical points. In particular the short range interaction implies
high collimated beams which results in low count rates and
moreover to increase sensitivity challenging beam slowing
techniques are necessary.

We are anyway confident that, being atom optics a fast developing
field, in the near future most of the experimental difficult
points will be clarified and atom interferometry will contribute
to the tests of non-newtonian gravitation.

\section{acknowledgements}
We kindly thank J.Vigu\'e for encouragement and many fruitful
discussions.



\newpage

\begin{figure}

\caption[Interferometer scheme]{\label{interferometer} General scheme of a separated arms interferometer of overall length $2 \times L$ and maximum separation $e$. $S$ and $D$ represent respectively the source and the detector. $G_1$, $G_2$ and $G_3$ are the beam splitters that divide and recombine the atomic beam. The wire, of diameter $d$ is set parallel and at the distance $d$ of the first part of the lower arm. As an illustrative example \cite{Ber} pp. 1-83, a three gratings Mach-Zehnder interferometer of D. Pritchard's group, MIT, uses nanofabricated gratings of period $\Lambda=200nm$ separated by $L=0.6m$. The source is a rare gaz seeded supersonic beam of sodium. Their de Broglie wavelength $\lambda_{dB}=\frac{h}{Mv}\sim 16pm$. It corresponds to a diffraction angle $\theta=\lambda_{dB}/\Lambda$ about $80\mu rad$. The beam spacing $e\sim 50\mu m$ is big enough so that a septum can be inserted between the two arms. The signal is then recorded on an hot wire detector.}

\end{figure}

\begin{figure}

\caption[Comparison of results]{\label{comparison} Comparison of results presented in Ref. \cite{HSH} (heavy lines) and detection limits of an atom
interferometer assuming a Yukawa potential (thin line) or $n$
extra dimensions $n=1$ (dash), $n=2$ (dot) or $n=3$ (dot dash).
The vertical axis, generically labeled $|\alpha|$, represents
either $\alpha_Y$ either $\alpha_{1,2,3}$ depending on the chosen
scenario. Results in \cite{HSH} assume a Yukawa extra potential.
The upper and lower parts correspond respectively to different
wire diameters and beam/wire distances $d=100\mu m$ and $d=10\mu
m$. In the former case, the results for extra dimensions models
have been extrapolated to the Yukawa model by a dotted line in the
region $d \geq \lambda$.}

\end{figure}



\begin{references}
\bibitem{newton2D} This plane geometry has nevertheless the
advantage that the newtonian potential is constant in space giving
thus a null contribution to the phase difference whatever the
thickness of the plate.
\bibitem{specialcases} In fact, for integer $n$, ${}_2F_1(\frac{1}{2},
\frac{1+n}{2},\frac{3}{2},-x^2)$ has simple analytical expression
as it can be verified integrating directly the potential by
recurrence. See \cite{Grad} Eq. $2.271.6$ and Eq. $2.148.4$:\\
$\int
dx\;(1+x^2)^{-\frac{2n+1}{2}}=\sum\limits_{k=0}^{n-1}\frac{(-1)^k}{2k+1}{n-1\choose
k }\frac{x^{2k+1}}{(1+x^2)^{\frac{2k+1}{2}}}$\\ $\int
dx\;(1+x^2)^{-n}=\frac{x}{2n-1}\sum\limits_{k=1}^{n-1}\frac{(2n-1)(2n-3)...(2n-2k+1)}
{(n-1)(n-2)...(n-k)(1+x^2)^{n-k}}+\frac{(2n-3)!!}{2^{n-1}(n-1)!}\arctan
x$
\bibitem{fullnewton} Here again, exact analytical expressions taking into
account the finite length of the wire exist. We get
$V_{newton}^{\scriptscriptstyle
1D}(x,z)=2GM\mu[\arcsinh\frac{L}{2d}+\arcsinh\frac{z}{d+x}-\arcsinh\frac{L+z}{d+x}]$.
The corresponding phases can then be easily obtained from computer
algebra but are so intricate that only the limiting case presented
above seem to have practical interest.
\end{references}
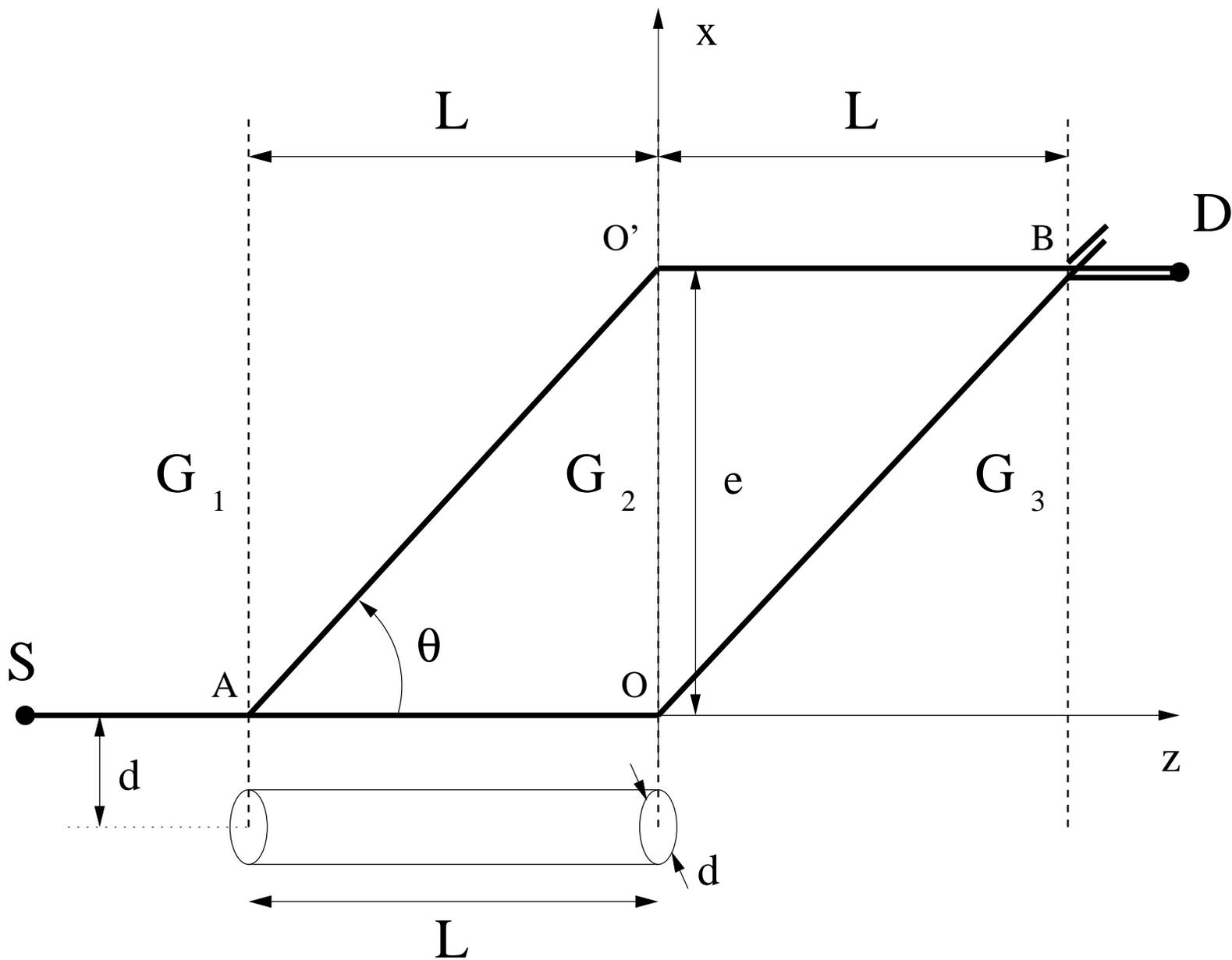
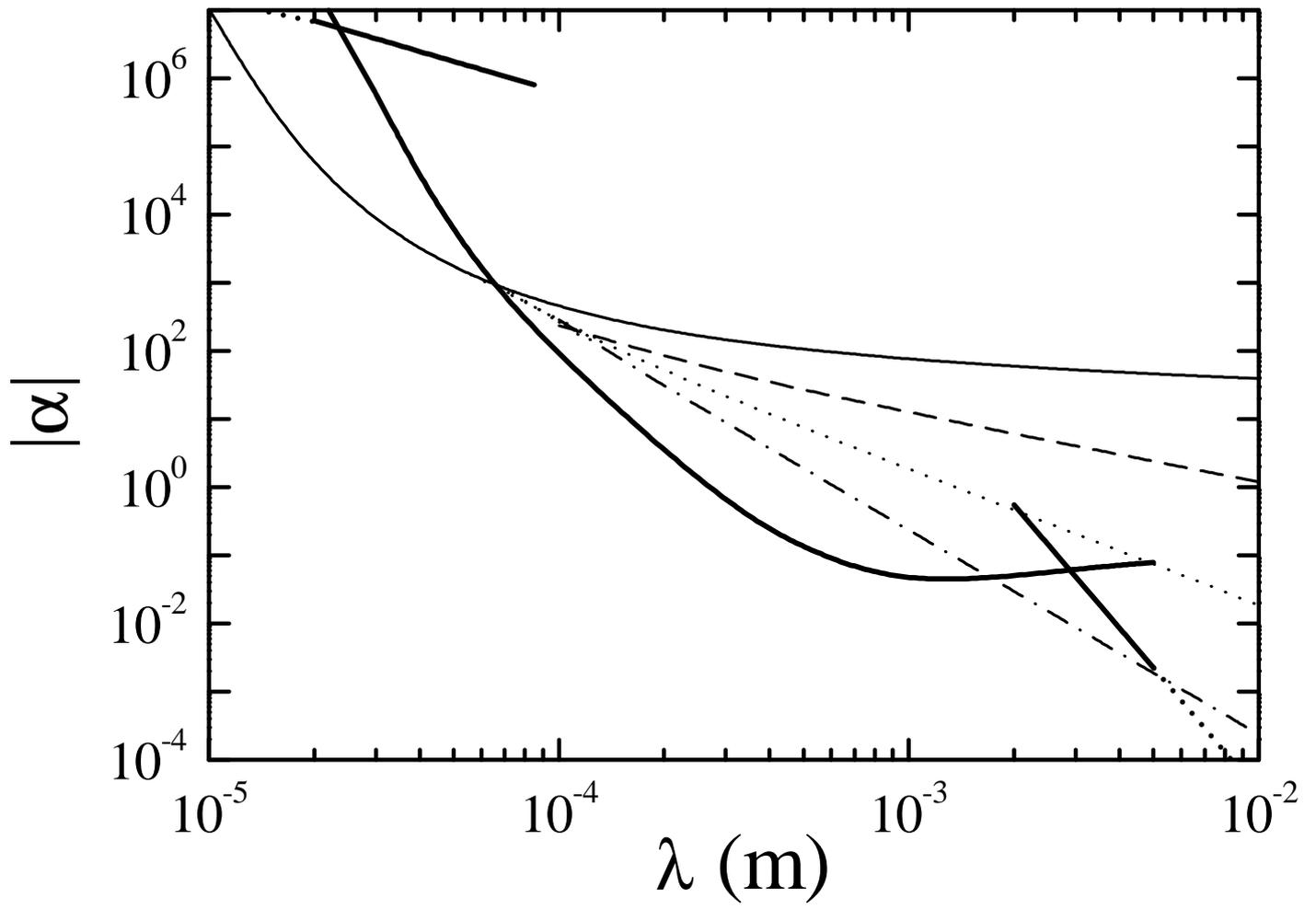
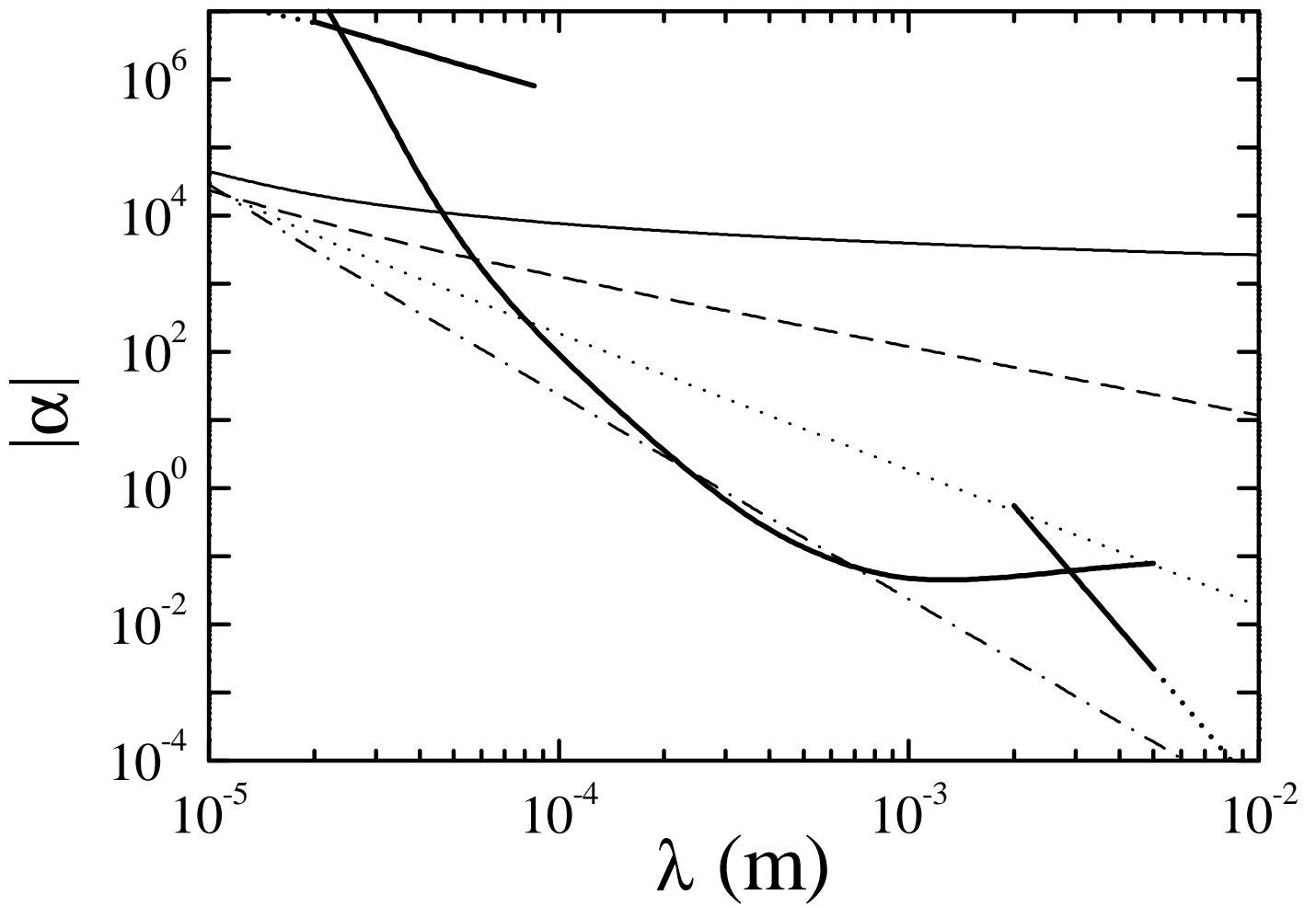
\end{document}